# Social Vehicle Swarms: A Novel Perspective on Social-aware Vehicular Communication Architecture

Yue Zhang, Fang Tian, Bin Song* *Member, IEEE,* and Xiaojiang Du, *Senior Member, IEEE*

*Abstract*—Internet of vehicles is a promising area related to D2D communication and internet of things. We present a novel perspective for vehicular communications, social vehicle swarms, to study and analyze socially aware internet of vehicles with the assistance of an agent-based model intended to reveal hidden patterns behind superficial data. After discussing its components, namely its agents, environments, and rules, we introduce supportive technology and methods, deep reinforcement learning, privacy preserving data mining and sub-cloud computing, in order to detect the most significant and interesting information for each individual effectively, which is the key desire. Finally, several relevant research topics and challenges are discussed.

## I. Introduction

INTERNET of vehicles (IoV) is a particular case, with vehicles being basic units, of internet of things (IoT) [1], which allows objects or devices to interact and communicate, indicating that an intrinsic component of IoT or IoV is device-to-device (D2D) communication [2]. IoV aims to build an intelligent system to improve the quality of driving or living; formally, to increase the quality of experience (QoE) or quality of service (QoS) [3]. Further study of IoV could lead to integration with smart cities, where each building, house, or even each individual device is capable of communication via wired or wireless access.

On the other hand, online social networks (OSNs) have gained a growing amount of attention during recent years, and their use has almost become a daily necessity. Therefore, social characteristics, such as community or centrality, have been exploited to solve D2D communication problems [4], or to be applied to certain devices, or vehicles in this case, which may describe and express the relationship and intrinsic social patterns of users. Thus social-aware internet of vehicles (SIoV) emerges [5], which intends to connect social networks with real vehicles, or to construct a vehicular social network (VSN), so that a fast and efficient information transmission system between internet and real world is established to benefit both terminals.

Our concern is based on the idea of swarms, which are commonly applied to inserts or animals of similar size such as ants, fish, or birds. Assembled with swarm intelligence, these simple and frail individuals attain a significant potential for intelligent global behavior, which is inaccessible by any individual agent. Therefore, it is prone to imagine the capability of social vehicle swarms, superior to biologic swarms, functionally and intelligently. Furthermore, combining social characteristics with vehicles makes it more practical to study hidden patterns by analyzing the performances of relevant devices. In other words, the behavior of vehicles could imply the properties of users and their relationships. In order to handle swarms, we introduce agent-based modelling to integrate information and reveal patterns based on the behavior of each individual, which may lead to understanding hidden habits or invisible knowledge. Fig. 1 presents a scene of social vehicle swarms, along with relationships between its components, which will be further discussed later.

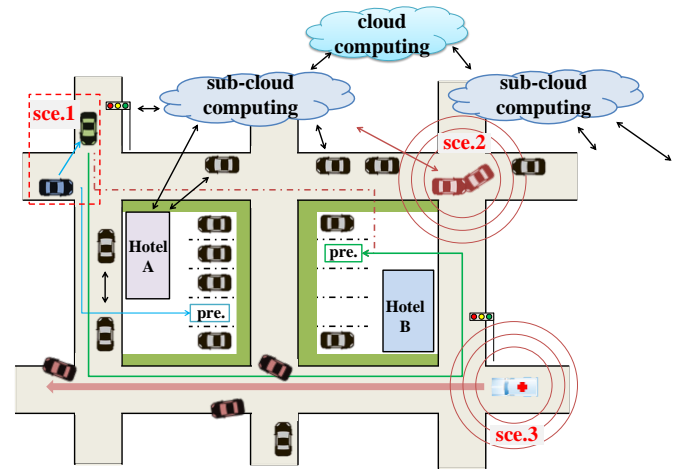

Fig. 1: One example of social vehicle swarms. (Pre. is the abbreviation of preservation.) Three particular scenarios (sce.1 ~ sce.3) will be discussed later.

This work has been supported by the National Natural Science Foundation of China (Nos.61271173 and 61372068), the Research Fund for the Doctoral Program of Higher Education of China (No.20130203110005), the Fundamental Research Funds for the Central Universities (No.K5051301033), the 111 Project(No. B08038), and also supported by the ISN State Key Laboratory.

Y. Zhang, F. Tian and B. Song are with the State Key Laboratory of Integrated Services Networks, Xidian University, 710071, China (e-mail: y.zhang@stu.xidian.edu.cn, fanglindatian@gmail.com, bsong@mail.xidian.edu.cn).
X. Du is with Dept. of Computer and Information Sciences, Temple University, Philadelphia PA, 19122, USA (email: dxj@ieee.org).

Under this model, the key problem is how to find the most important and interesting information efficiently and safely. Its difficulty lays in the vague definitions of importance and interest, since they vary between individuals and across time. Meanwhile, the growing velocity, volume and variety of data, brought on by the age of big data, increase the complexity and effort of such a task [6].

Security, on the other hand, is always an unavoidable and critical problem during the data mining process. Privacy preserving data mining (PPDM) offers a framework that can be regarded as an enhanced version of the traditional big data structure with minimization of risk [7]. It is common knowledge that conserving data leads to high security at the price of an unsatisfactory user experience. Conversely, over utilizing data increases the satisfaction but may jeopardize data providers' security. Consequently, a tradeoff is required to be met between these two extreme scenarios.

Customization is always an unavoidable yet interesting problem when maximization of the QoE of each individual, instead of averaged QoE, is considered. Thus the intersection of deep learning and reinforcement learning, deep reinforcement learning (DRL), seems to be a reliable solution. It aims to discover hidden patterns behind these data, to recognize certain objects, and to achieve a quality of experience that is customizable for each individual, if provided with a large quantity of data. Mnih et al. has provided one instance to demonstrate DRL's great potential for complicated policies [8].

In the cyber layer, cloud computing is already famous in a variety of research and application fields. It stands in a high altitude and arranges all available and configurable resources. Sub-cloud computing, such as fog or edge computing, however, is located closer to the terminals. It may be associated with cloud computing or work individually. In a sense, sub-cloud computing is much more flexible and efficient than cloud computing [9].

The main contribution of this paper is summarized as follows:
- We propose a novel perspective, social vehicle swarms, to regard vehicles as the main agents in swarms, being the first time to the best of our knowledge.
- We apply an agent-based model to handle social vehicle swarms and analyze its components. This model mirrors the inherent social characteristics and patterns into superficial vehicles and devices, making them more accessible to be studied and applied.
- Under the foundation of agent-based model, several supportive methods are introduced and embedded for the purpose to maximize the QoE of each individual.

The following section presents the architectures of social vehicle swarms, and corresponding technology and methods are discussed in Section III. Section IV discusses future research directions and challenges and Section V concludes our argument.

## II. SOCIAL VEHICLE SWARMS ARCHITECTURE

The mystery of biologic swarms remains attractive. Basically, biological researchers apply chips on some individuals within swarms or record data of a particular location for a period of time in order to gather data and to study swarm intelligence. Swarm intelligence is the core for biologic swarms. The emergence of global intelligence significantly improves the capability of each frail individual. Similarly, it is probable that social vehicle swarms might follow the same pattern and lead to a higher QoE for humans. Therefore, the analysis, study, and establishment of this social vehicle swarm system are worthwhile.

Fig. 2 demonstrates social vehicle swarm architecture. Data are generated by users and sensors, establishing a big data environment. Data are either stored locally or uploaded to sub-cloud computing terminals, based on different functions. Customization systems are designed and hidden patterns are discovered with the assistance of DRL, where PPDM ensures safety. Finally, QoE of each individual is satisfied. From the perspective of agent-based modeling, the elements that require careful treatment are agents, environments and rules. Thus, details of these elements are presented as follows.

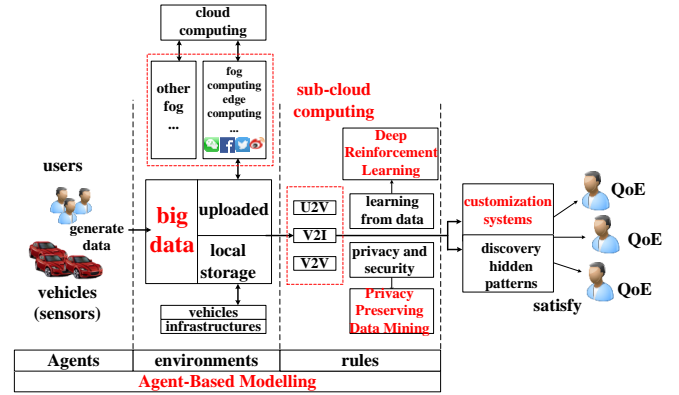

Fig. 2: The architecture of social vehicle swarms

### A. Agents

Agents are subjects in agent-based modelling, in this case, individuals in swarms. These agents are supposed to reveal swarm intelligence if they interact with certain environments or other agents based on well-designed rules. Agents in social vehicle swarms are straightforward and simple, similar to the agents in IoV, namely users, vehicles, and infrastructures. Notice that infrastructures can also be considered a component of environments that interact with users and vehicles.

### B. Environments

Environments are actually big data environments, which are relatively complicated when dynamic and static information is taken into account. They involve online social networks, which can vary rapidly, as well as structures, which can remain in a particular state for a period of time. Therefore, data management is a critical aspect in agent-based modelling.

The establishment of environments is also considered in cyber layers. Cloud computing or sub-cloud computing is utilized to improve the efficiency of data transmission and application. Wired with OSNs, environments are much more complicated and dynamic, with the capability to spread data without location constraints, the lack of which is a serious drawback for biologic environments.

## C. Rules

Rules for this model may be the most vital and difficult, which require careful and patient treatments. Two classes of rules are considered: rules for agents alone, and rules for their communications.

Individual rules are relatively trivial. Physical conditions, such as velocity or the communicative range of each agent are defined by certain physical parameters. Thus, agents play different roles under different circumstances. Users are data providers, collectors, and decision makers. Vehicles are executors controlled by users, whereas infrastructures act as databases, which transmit and store valuable and useful information and provide it to users when required.

More attention is paid to communicative rules, which focus on connections between pairs of components in our model, namely user-to-vehicle (U2V), vehicle-to-infrastructure (V2I), and vehicle-to-vehicle (V2V) communications, to realize swarm behaviors. Therefore, we examine details of each communication pair meticulously without revealing a complete set of rules, which is impractical due to the complexity of social vehicle swarm behavior.

*1) User-to-vehicle*

Conservative interactions between users and vehicles are widely understood. Higher intelligence, however, is capable of providing a more convenient and satisfactory experience to users. One of the most significant aspects of such communication is identification of the current driver, which could improve security and quality of driving. Data of identified driver on driving habit may be accessed, if he is not denied by the trust list. These data may be stored in the vehicle, since its condition is almost stable, or in OSNs, which vary more frequently. After the data fusion procedure, assistant or automatic driving takes charge. This customizable system processes advantages for taking variation among drivers into consideration, which leads to less ambiguity in controlling the system.

The capability of dealing with disorder is critical. Once an accident occurs, several procedures, such as first aid or contact with hospitals, are requested. Computers seem to be more reliable for such tasks, since they are not emotional or anxious at an urgent situation and can probably remain reasonable enough to make the optimal choice based on current conditions.

*2) Vehicle-to-infrastructure*

Interactions between swarms and environments are critical in this model, because they express swarm behavior and swarm intelligence. Certain infrastructures can generate and store local data to provide to neighboring agents since closeness may infer importance. Meanwhile, some infrastructures may upload data to OSNs to support distant agents.

An intelligent parking system may be a promising research topic. Parking is becoming a growing problem with the increasing number of vehicles. A common phenomenon is the failure to find parking spaces. This may discourage benefits for both demanders and suppliers, such as customers and owners of restaurants. An improvement focuses on resource allocation, along with preservation of parking places, which takes the behaviors of other vehicles into consideration. As Scenario 1 in Fig. 1, Hotel A was the best choice for both vehicles, Blue and Green. However, Vehicle Green is recommended Hotel B, which is a suboptimal choice for him, because Blue navigates and preserves ahead and Hotel A is saturated. The failure of coordination wastes resources and discourages QoE.

Another aspect for vehicles is self-checking, which will report relevant details of any device or system malfunction. An advanced system may also analyze images generated from cameras monitoring surroundings. With an appropriate arrangement of cameras, a potential danger or undesired situation around vehicles may be detected. When social networks are involved, however, data from neighboring vehicles may provide a more comprehensive understanding of environmental conditions. Typically, this requires shared access to data from a stranger vehicle. Therefore, methods for privacy are considered.

*3) Vehicle-to-vehicle*

Vehicle-to-vehicle communications are the core of social vehicle swarms, which is reasonable since interactions between agents in natural systems play a similar role. Superficially, each agent expects to follow its own route and to maximize its own outcomes. Intrinsically, however, all agents follow certain patterns unconsciously. Discovery such patterns may formulate a plan, evaluate a candidate policy, and finally improve the quality of experience.

One significant aspect of both vehicular and biologic swarms is navigation, since it is the initiative curiosity. Navigation devices or software are familiar, but most of current examples plan routes offline because they are unable to take other vehicles' data into consideration. One strict requirement of a reliable navigation system is communication efficiency, which means that once an unexpected scenario occurs, such as a car accident, presented as Scenario 2 in Fig. 1, eligible vehicles or infrastructures are obligated to spread this information for the benefit of other vehicles' route planning.

Vehicle control is also a popular topic, which focuses on assisted and automatic driving. They may liberate drivers to some degree without jeopardizing anything. This requires a combination of a variety of techniques such as sensors or fuzzy control systems. Some systems or vehicles are realized in an experimental scenario. They, however, still need to be improved in order to satisfy the real application.

An intelligent vehicle control system takes V2V into consideration. Car accidents, for instance, are mainly caused by failure to detect dangerous vehicles or obstacles. This might be improved if signals are transmitted through vehicles instead of being dependent on human observation.

Another improvement locates the priority of street utilization. Emergency vehicles should be guaranteed higher priority. Consider an ambulance in route to an emergency. With the assistance of V2V communication, knowledge of an emergency is spread through OSNs to all vehicles that impede the ambulance's route, and they are manipulated to make way to ensure its smooth travel, presented as Scenario 3 in Fig. 1.

Notice that the priority is not designed or fixed. For instance, a taxi with a patient is more urgent than an inactive ambulance. This significance may be discovered through social networks if someone posted it online. Then with certification of the particular vehicle's emergency, it could obtain similar priority to that of an active ambulance.

With highly developed vehicle and traffic control technology, drivers could enjoy themselves, similar to a passenger, without suffering from driving problem, which satisfies QoE in some degree.

## III. METHODS AND TECHNOLOGY

Based on the agent-based model framework, certain methods and technology are involved to solve our key problem, which is to discover the most important and useful information within an acceptable period. This might be difficult since importance and usability are vaguely defined and vary among individuals, and because privacy is always jeopardized unconsciously. Under the big data circumstances, various data mining methods are developed for certain applications. Here we present only a few of them, which seem to be the most promising ones for our model.

### A. Big data and privacy preserving data mining

The ecosystem of privacy preserving data mining or traditional big data consists of four agents, namely the data provider, the data collector, the data miner and the decision maker, as presented in Fig. 3, which details the relevant components in Fig. 1. One distinguishing and interesting feature of social vehicle swarms is that data are generated from two distinct fields. One field is vehicle sensors, such as cameras, whereas the other originates from the internet, specifically social networks. These data differ in forms and volumes. Data from sensors are relatively stable and the number of sensors on a specific vehicle is typically fixed. Therefore, it is not difficult to calculate the constant, such as the amount of data from that vehicle. The other source, social networks, however, creates data with more awareness. They involve subjective motion and opinion along with objective images or videos which may not appear to be valuable. These data, however, may be critical and conclusive for others. Thus, the ability to discover such valuable information is vital.

One approach to protecting privacy during this process is to limit the access, which means that data collectors can only be served with data passively, rather than actively seek data. This operation may block some suspicious procedures and reduce the risk. An improved version might focus on establishing hierarchical accesses for different collectors. A less creditable collector may only be able to gather data in a safer layer whereas a trustworthy one could access more sensitive information.

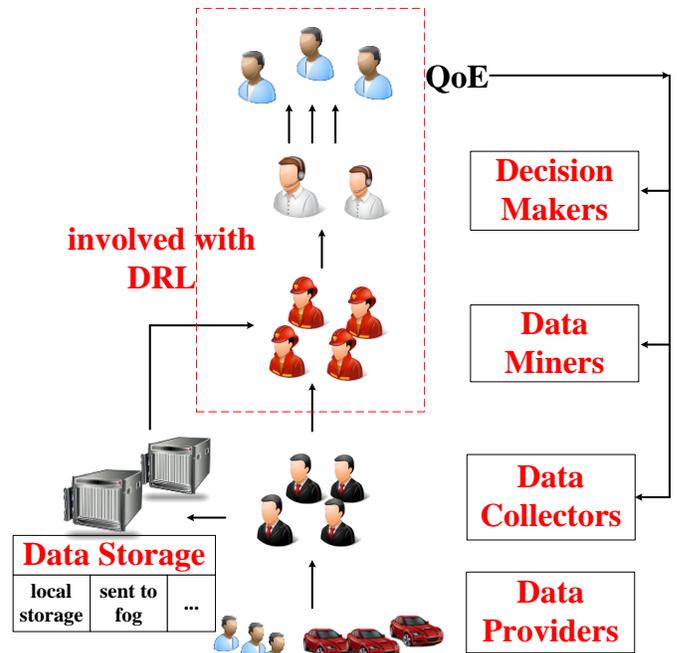

Fig. 3: The ecosystem of PPDM

Data collectors handle pre-processing in preparation for further procedures. Data with high similarities are integrated and compressed so that they can be stored or analyzed conveniently and efficiently. After the integration process, a procedure of cleansing is designed to eliminate problematic data, which are missing or inaccurate. This cleansing process also feeds back in order to improve the performance of previous ones. Specifically, it may assist to filter and prevent the unacceptable data to increase the efficiency of collectors, pushing it towards the possibility of collecting social information in real time. For the purpose of privacy protection, anonymity is often integrated into this process. The latter processes, such as data miners are mainly interested in the patterns or trends from data. The ownership of each specific piece of information is likely unimportant. Therefore, anonymization operations hide sensitive parts, such as the source's identity, and retain valuable segments.

Data storage and management mainly concerns computer scientists. Data are ultimately sent to data miners in two types of paradigms, streaming and batch. Streaming data require almost no storage. They are transmitted directly to be analyzed and will be deposited after being studied. The value of streaming data is related to its freshness, therefore widely applied for online algorithms. Batch data are stored and may be stored again after being utilized.

Data miners are the core of this chain in which hidden informative patterns are discovered behind superficial phenomena. This stage is wired with DRL, which will be discussed later. In terms of social vehicle swarms, data miners attempt to discover the social patterns, such as rush hour in a city, and deliver them to decision makers for the further process.

Finally, when miners discern the patterns or the categories from the data, decision makers are obligated to make an optimal choice, which may be designed for individuals, such as suggestions of an optimal route for drivers. Then, feedback from users is sent back to collectors, miners, and decision makers to improve QoE.

Privacy is difficult for miners and decision makers, because personal data are required for customization purposes. Otherwise each individual would receive identical information and suggestions, which makes no difference from current advertisement outcomes. On the other hand, if customization is achieved, which means personal data are involved, then safety and security are not guaranteed. One possible solution is the application of fake identification to mask the true user behind the provided information. This can mitigate the risk to some extent, but cannot eliminate it entirely.

One drawback of this paradigm is indiscrimination, which could not satisfy the requirements of customization. Therefore, we further introduce DRL to fix this problem, based on the relatively private and secure environment provided by PPDM.

### B. Deep reinforcement learning

Deep reinforcement learning is a novel algorithm, which combines two outstanding methods in machine learning, namely, deep learning and reinforcement learning. Deep learning is a neural network frame constructed with multiple hidden layers. Several examples have demonstrated its capability of reaching a noteworthy accuracy in tasks such as image recognition. This accuracy is achieved only if the deep neural networks are fed with plenty of data, which fits nicely with the phenomenon of big data.

On the other hand, reinforcement learning is a method that provides the possibility of an agent discovering dominating policy with certain rewards. Thus, the optimal choice is obtained by maximizing the rewards of an agent. Notice that this optimization problem varies among all individuals, which motivates the idea of customization in social vehicle swarms. One distinguishing feature of reinforcement learning is the tradeoff between exploitation and exploration, which offers it the capability to keep discovering new policies while simultaneously ensuring agent payoffs.

Equipped with advantages of these two aspects, DRL seems to be a promising solution to establish customization systems under the circumstance of big data. Traditional neural networks are adapted as a deep Q-network (DQN), which locates the weights of neurons into Q-values from reinforcement learning.

Another reason to apply DRL to social vehicle swarms is because socially aware networking alone is not capable of social learning in order to analyze data or to recognize patterns [10]. Therefore the introduction of DRL could fix the drawback.

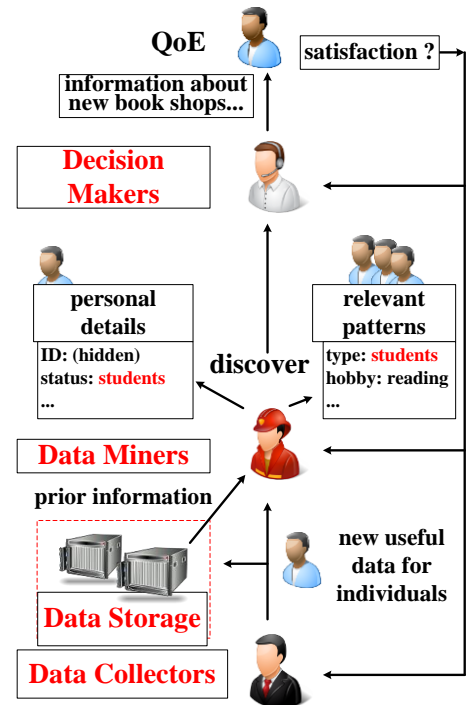

Fig.4: A demonstration of structure of deep reinforcement learning

Applying DRL to agent-based model can be considered from two distinct perspectives, macrocosmic and microcosmic. Macrocosmic view regards the whole paradigm as a recommendation system, which takes information of users as input and learns to maximize their QoE based on their preferences [11]. Fig. 4 presents a quick example for DRL, for recommending new book shops to a particular user based on him being a student given the pattern of students, even though his ID is hidden, for PPDM being involved. On the other hand, each role in Fig. 4 is not monopolistic, making a competition model or a game model abstractly. Therefore, it can be solved by applying multi-agent systems with DRL to each agent in such environments [12]. Similar to a free market, each unit manages to maximize his payoffs by interacting with others, ending up equilibrium.

One problem that DRL faces is the dilemma of efficiency and accuracy. With the assistance of GPU or cloud computing and after a certain period of training, DRL could reach a remarkable accuracy. This, however, could not satisfy a rapid demand. Therefore we introduce sub-cloud computing, which sacrifices accuracy to some extent for efficiency, constructing a hybrid structure.

### C. Sub-cloud computing

The advantages of cloud computing are well-known, yet it is unpractical for scenarios demanding a rapid response. Sub-cloud computing, which is an overall concept referring to fog computing, edge computing, and other similar technologies, is more flexible in social vehicle swarms. It can be regarded as a version of cloud computing which is closer to terminals; in this case, to all vehicular agents. Sub-cloud computing provides a partial overall control of the data interaction. In other words, it allocates data to collectors synthetically according to their preference, increasing the efficiency of data, adding the probability of real time communication. Sub-cloud computing may be linked to cloud computing, but they also can be regarded in different responsibilities, as in Fig. 5.

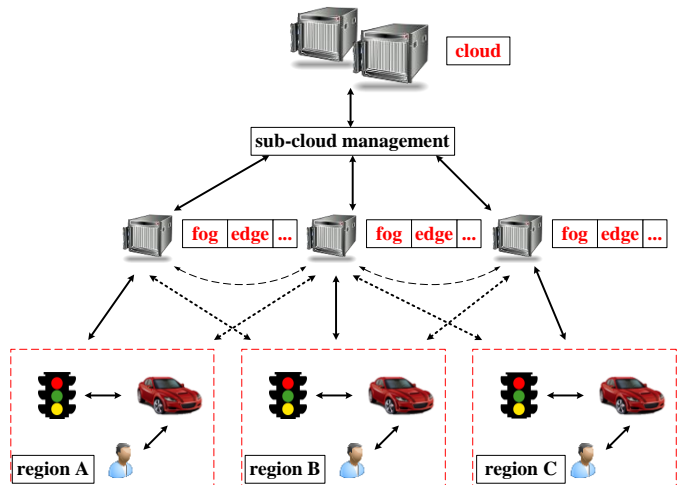

Fig. 5: Structure of sub-cloud computing in social vehicle swarms

Sub-cloud management manipulates communications between sub-clouds and clouds. Data related only to an event with low importance may be stored and processed in sub-cloud layer without the necessary of bothering cloud layer. Meanwhile, the capabilities of cloud computing are not abandoned. Thus, the combination of cloud computing and terminals are introduced. Low-effort work can be solved at terminals, whereas difficult problems are transmitted to cloud.

## IV. FUTURE RESEARCH DIRECTIONS AND CHALLENGES

Although a variety of topics with high similarity to IoV are widely discussed, social vehicle swarms constitutes a novel perspective and a new direction for such research. Complexity may be the most crucial challenge for social vehicle swarms. Firstly, as mentioned above, details of the communicative rules require careful treatment. If multiple candidates require communication, for instance, the acceptance strategy ought to maximize users' QoE given limited channel capacity. Secondly, vague concepts such as importance or usefulness are difficult to define. DRL is capable of dealing with such uncertainty. The process of training DRL networks, however, is a problem, since it requires a considerable quantity of data. Lastly, it is always difficult to ensure safety because loopholes are popular in any advanced system or product. Meanwhile, high technologies are always used for unacceptable or illegal purposes. Thus, caution should always be emphasized.

Besides these challenges, we present relevant research directions where social vehicle swarms could be applied, along with new challenges.

### A. Internet of multimedia things

Internet of multimedia things (IoMT) is a novel paradigm and a particular subset of IoT, allowing the interaction and communication among multimedia devices [13]. With the cooperation of multimedia devices and services, certain applications, such as telemedicine, could be realized. QoS or QoE is also considered in this paradigm, sharing the high similarity with our social vehicle swarms.

However, one of open issues in IoMT is the requirement of customization, which is sophisticated discussed in this paper. Information and knowledge can be sensed and the responsive signals, which are previously learned, can be obtained, with the assistance of deep reinforcement learning. Therefore, the research and study on either topic may have a positive impact on the other.

### B. Home automation

Home automation focuses on the intelligence of our daily life, regarding vehicles as one of its devices [14]. Intuitively, a home automation system combines information from the internet, devices, and indoor sensors to control certain devices, such as an air conditioner, or to offer advice, such as health suggestions. The development of home automation paves the way for intelligent cities, since houses or departments are unavoidable elements in a city.

One critical problem with combining social vehicle swarms and home automation is data synchronization, which targets data sharing between home and vehicles. Vehicles generate amounts of data during the day, and destination of such data requires circumspection to retain useful data and erase the rest. Notice that customization systems are also required to determine the value of data. Some data may be potentially valuable in a long term instead of being useful immediately, which further complicates such a task.

### C. Urban computing

Urban computing aims to solve urban problems or to estimate urban policy by analyzing data from all available resources. Basic achievements are air or noise condition estimations, with data gathered directly from sensors or indirectly by indication and inferences from relevant information.

Zheng has proposed an extraordinary method of determining gas consumption with data from the GPS trajectories of taxis [15]. If more personal vehicles and data are involved, constructing swarms of vehicles, the results may be more accurate, and more interesting results will likely be revealed.

However, urban computing takes more objects, such as points of interest (POIs), into consideration, and methods such as tensor decomposition are involved. Therefore, the gap between urban computing and social vehicle swarms is worth noticing, although both aspects target similar problems in urban environments and intend to improve living quality.

## V. CONCLUSION

This paper has introduced a novel perspective, social vehicle swarms, which provides an agent-based model to research and analyze social-aware internet of vehicles. Agents, environments and rules are carefully studied and analyzed. We also have discussed technologies, privacy preserving data mining, deep reinforcement learning, and sub-cloud computing, behind this model, which are critical for social vehicle swarms. Finally, some future research directions and challenges have been considered. Hopefully this model could discover hidden patterns behind our daily lives, and establish an intelligent system for vehicles and cities based on social networks to improve the living conditions and quality of experience.